\documentclass[aps,prd,superscriptaddress,preprint,tightenlines,nofootinbib,floatfix]{revtex4}

\usepackage{graphicx} 
\usepackage{dcolumn}  
\usepackage{bm}       

\usepackage{latexsym}
\usepackage{amsmath}
\usepackage{xspace}

\begin{document}




\preprint{CLNS 09/2062}  
\preprint{CLEO 09-15~~~}    



\title{Measurement of absolute branching fractions of inclusive semileptonic decays of charm and charmed-strange mesons}


%


\author{D.~M.~Asner}
\author{K.~W.~Edwards}
\author{J.~Reed}
\author{A.~N.~Robichaud}
\author{G.~Tatishvili}
\author{E.~J.~White}
\affiliation{Carleton University, Ottawa, Ontario, Canada K1S 5B6}
\author{R.~A.~Briere}
\author{H.~Vogel}
\affiliation{Carnegie Mellon University, Pittsburgh, Pennsylvania 15213, USA}
\author{P.~U.~E.~Onyisi}
\author{J.~L.~Rosner}
\affiliation{University of Chicago, Chicago, Illinois 60637, USA}
\author{J.~P.~Alexander}
\author{D.~G.~Cassel}
\author{S.~Das}
\author{R.~Ehrlich}
\author{L.~Fields}
\author{L.~Gibbons}
\author{S.~W.~Gray}
\author{D.~L.~Hartill}
\author{B.~K.~Heltsley}
\author{J.~M.~Hunt}
\author{D.~L.~Kreinick}
\author{V.~E.~Kuznetsov}
\author{J.~Ledoux}
\author{J.~R.~Patterson}
\author{D.~Peterson}
\author{D.~Riley}
\author{A.~Ryd}
\author{A.~J.~Sadoff}
\author{X.~Shi}
\author{S.~Stroiney}
\author{W.~M.~Sun}
\affiliation{Cornell University, Ithaca, New York 14853, USA}
\author{J.~Yelton}
\affiliation{University of Florida, Gainesville, Florida 32611, USA}
\author{P.~Rubin}
\affiliation{George Mason University, Fairfax, Virginia 22030, USA}
\author{N.~Lowrey}
\author{S.~Mehrabyan}
\author{M.~Selen}
\author{J.~Wiss}
\affiliation{University of Illinois, Urbana-Champaign, Illinois 61801, USA}
\author{M.~Kornicer}
\author{R.~E.~Mitchell}
\author{M.~R.~Shepherd}
\author{C.~M.~Tarbert}
\affiliation{Indiana University, Bloomington, Indiana 47405, USA }
\author{D.~Besson}
\affiliation{University of Kansas, Lawrence, Kansas 66045, USA}
\author{T.~K.~Pedlar}
\author{J.~Xavier}
\affiliation{Luther College, Decorah, Iowa 52101, USA}
\author{D.~Cronin-Hennessy}
\author{K.~Y.~Gao}
\author{J.~Hietala}
\author{R.~Poling}
\author{P.~Zweber}
\affiliation{University of Minnesota, Minneapolis, Minnesota 55455, USA}
\author{S.~Dobbs}
\author{Z.~Metreveli}
\author{K.~K.~Seth}
\author{B.~J.~Y.~Tan}
\author{A.~Tomaradze}
\affiliation{Northwestern University, Evanston, Illinois 60208, USA}
\author{S.~Brisbane}
\author{J.~Libby}
\author{L.~Martin}
\author{A.~Powell}
\author{P.~Spradlin}
\author{G.~Wilkinson}
\affiliation{University of Oxford, Oxford OX1 3RH, UK}
\author{H.~Mendez}
\affiliation{University of Puerto Rico, Mayaguez, Puerto Rico 00681}
\author{J.~Y.~Ge}
\author{D.~H.~Miller}
\author{I.~P.~J.~Shipsey}
\author{B.~Xin}
\affiliation{Purdue University, West Lafayette, Indiana 47907, USA}
\author{G.~S.~Adams}
\author{D.~Hu}
\author{B.~Moziak}
\author{J.~Napolitano}
\affiliation{Rensselaer Polytechnic Institute, Troy, New York 12180, USA}
\author{K.~M.~Ecklund}
\affiliation{Rice University, Houston, Texas 77005, USA}
\author{J.~Insler}
\author{H.~Muramatsu}
\author{C.~S.~Park}
\author{E.~H.~Thorndike}
\author{F.~Yang}
\affiliation{University of Rochester, Rochester, New York 14627, USA}
\author{S.~Ricciardi}
\affiliation{STFC Rutherford Appleton Laboratory, Chilton, Didcot, Oxfordshire, OX11 0QX, UK}
\author{C.~Thomas}
\affiliation{University of Oxford, Oxford OX1 3RH, UK}
\affiliation{STFC Rutherford Appleton Laboratory, Chilton, Didcot, Oxfordshire, OX11 0QX, UK}
\author{M.~Artuso}
\author{S.~Blusk}
\author{S.~Khalil}
\author{R.~Mountain}
\author{K.~Randrianarivony}
\author{T.~Skwarnicki}
\author{S.~Stone}
\author{J.~C.~Wang}
\author{L.~M.~Zhang}
\affiliation{Syracuse University, Syracuse, New York 13244, USA}
\author{G.~Bonvicini}
\author{D.~Cinabro}
\author{A.~Lincoln}
\author{M.~J.~Smith}
\author{P.~Zhou}
\author{J.~Zhu}
\affiliation{Wayne State University, Detroit, Michigan 48202, USA}
\author{P.~Naik}
\author{J.~Rademacker}
\affiliation{University of Bristol, Bristol BS8 1TL, UK}
\collaboration{CLEO Collaboration}
\noaffiliation


%
\date{December 21, 2009}

\begin{abstract}
We have measured the inclusive semileptonic branching fractions of
$D^0$, $D^+$, and $D^+_s$ mesons.
For these measurements, we have used the full CLEO-c open-charm
data samples, $818$ pb$^{-1}$ at $E_\text{CM} = 3.774$ GeV, giving
$D^0 \bar{D}^0$ and $D^+ D^-$ events,
and $602$ pb$^{-1}$ at $E_\text{CM} = 4.170$ GeV, giving
$D^{\ast \pm}_s D^\mp_s$ events.
We obtain
$\mathcal{B}(D^0 \to X e^+ \nu_e) = (6.46 \pm 0.09 \pm 0.11)$\%,
$\mathcal{B}(D^+ \to X e^+ \nu_e) = (16.13 \pm 0.10 \pm 0.29)$\%,
and
$\mathcal{B}(D^+_s \to X e^+ \nu_e) = (6.52 \pm 0.39 \pm 0.15)$\%,
where the first uncertainties are statistical and the second are systematic.
From these and lifetimes obtained elsewhere,
we obtain the ratios of semileptonic decay widths
$\Gamma (D^+ \to X e^+ \nu_e) / \Gamma (D^0 \to X e^+ \nu_e) =
0.985 \pm 0.015 \pm 0.024$
and
$\Gamma (D^+_s \to X e^+ \nu_e) / \Gamma (D^0 \to X e^+ \nu_e) =
0.828 \pm 0.051 \pm 0.025$.
The ratio of $D^+$ and $D^0$ is consistent with the isospin symmetry
prediction of unity,
and the ratio of $D^+_s$ and $D^0$ differs from
unity, as expected.
\end{abstract}

\pacs{13.20.Fc}
\maketitle

%
\section{\label{sec:introduction}Introduction}

As part of the CLEO-c analyses of
exclusive~\cite{Coan:2005iu,Huang:2005iv,ichep06:ygao,Mitchell:2008kb,Besson:2009uv,:2009cm,Ecklund:2009fi}
and
inclusive semileptonic decays~\cite{Adam:2006nu},
this article presents measurements of
$D^0$, $D^+$, and $D^+_s$ inclusive semileptonic branching fractions
using the complete CLEO-c data sets.
Using these results and known lifetimes, we
also report the ratios of the widths
$\Gamma (D^+ \to X e^+ \nu_e) / \Gamma (D^0 \to X e^+ \nu_e)$
(which is expected to be unity due to isospin symmetry)
and $\Gamma (D^+_s \to X e^+ \nu_e) / \Gamma (D^0 \to X e^+ \nu_e)$
(which is not expected to be unity~\cite{Voloshin:2001xi,Scora:1995ty},
though with poor theoretical precision). These measurements are
important in their own right, and, due to similarities between the $D$
and $B$ sectors, will also improve understanding of $B$ semileptonic
decays.
In particular, knowledge of the previously unmeasured ratio
$\Gamma (D^+_s \to X e^+ \nu_e) / \Gamma (D^0 \to X e^+ \nu_e)$
enables a more reliable prediction of the difference of the
inclusive decay rates between $B^0$ and $B^+$ mesons in
$b \to u \ell^+ \nu_\ell$ decays,
thereby reducing theoretical uncertainty~\cite{Voloshin:2001xi}
in determination of weak mixing parameter $V_{ub}$.

%
%
Two sets of open-charm data samples are used to study the
semileptonic decays of charm and charmed-strange mesons.
In $e^+ e^-$ collisions provided by
the Cornell Electron Storage Ring (CESR),
the CLEO-c detector has collected integrated luminosities of
$818$ pb$^{-1}$
at the
center-of-mass energy $E_\text{CM} = 3.774$ GeV
near the peak of the $\psi(3770)$ resonance which decays to
$D\bar{D}$ pairs,
and
$602$ pb$^{-1}$
at $E_\text{CM} = 4.170$ GeV near the peak production of
$D^{\ast \pm}_s D^\mp_s$ pairs.
The former data set
contains $3.0 \times 10^{6}$
$D^0 \bar{D}^0$
and $2.4 \times 10^{6}$
$D^+D^-$ pairs,
and is used to study $D^0$ and $D^+$ semileptonic decays.
The latter data set
contains $0.6 \times 10^{6}$
$D^{\ast \pm}_s D^\mp_s$ pairs,
and is used to study $D^+_s$ semileptonic decays.
We have previously reported~\cite{Adam:2006nu} measurements of
inclusive semileptonic decay branching fractions of $D^0$
and $D^+$ mesons with a subsample of the former data set.

%
%
The remainder of this article is organized as follows.
The CLEO-c detector is described in Sec.~\ref{sec:detector}.
Event reconstruction and selection criteria are described
in Sec.~\ref{sec:event_selectuon}.
The analysis procedure to extract semileptonic
decay rates is covered in Sec.~\ref{sec:analysis}.
Results for inclusive spectra are presented in Sec.~\ref{sec:results}.
Systematic uncertainty in our measurements is evaluated in
Sec.~\ref{sec:systematic_uncertainty}.
Finally, in Sec.~\ref{sec:conclusion} a summary of our results is provided.

\section{\label{sec:detector}The CLEO-\lowercase{c} Detector}

%
%
The CLEO-c detector~\cite{Briere:2001rn,Kubota:1991ww,cleoiiidr,cleorich}
is a general-purpose solenoidal detector
equipped with
four concentric components:
a six-layer vertex drift chamber,
a 47-layer main drift chamber,
a ring-imaging Cherenkov (RICH) detector,
and a cesium iodide electromagnetic calorimeter.
The detector provides acceptance of $93$\% of the full $4 \pi$
solid angle for both charged particles and photons.
The main drift chamber provides specific-ionization ($dE/dx$)
measurements that discriminate between charged pions and kaons.
The RICH detector covers approximately $80$\% of $4 \pi$ and provides
additional separation of pions and kaons at high momentum ($\ge 700$ MeV).
Electron identification is based on a likelihood variable that combines
the information from the RICH detector, $dE/dx$,
and the ratio of electromagnetic shower energy to track momentum ($E/p$).
A \textsc{geant}-based~\cite{geant} Monte Carlo (MC) simulation
is used to study efficiencies of signal and background events.
Physics events are generated
by \textsc{evtgen}~\cite{evtgen}, tuned with improved knowledge of
charm decays~\cite{:2007zt,:2008cqa,:2009ni,pdg2008},
and
final-state radiation (FSR) is modeled by
\textsc{photos}~\cite{photos}.

%
%
\section{\label{sec:event_selectuon}Event Selection}

Charm or charmed-strange mesons are always
produced in pairs in our open-charm data samples.
Since the data are taken just above threshold,
the mesons are produced in
a very clean environment with no additional particles except,
in the case of the $D_s D^{\ast}_s$, a photon or a neutral pion
from the $D^{\ast}_s$ decay.
The analysis proceeds by
first defining a single tag (ST) sample,
in which one of the $D$ (or $D_s$) mesons in a
$D \bar{D}$ (or $D_s D^{\ast}_s$) event is reconstructed in a chosen
hadronic decay mode, and a further double tag (DT) subsample
in which an additional recoiling electron (or positron)
is required as a signature
of the signal semileptonic decay.
Absolute semileptonic branching fractions for charm or charmed-strange
mesons can then be obtained
from the fraction of the ST sample that is DT, without
requiring any knowledge of the integrated luminosity or how many mesons
are produced.

\subsection{\label{sec:event_selectuon::tag_selection}Tag Selection}

To minimize the combinatorial
backgrounds and systematic uncertainties, three very clean
tag modes composed of only charged particles
are used:
$\bar{D}^0 \to K^+ \pi^-$,
$D^- \to K^+ \pi^- \pi^-$,
and
$D^-_s \to \phi \pi^-$.
Here, the notation
$D^-_s \to \phi \pi^-$ is a shorthand label for
$D^-_s \to K^- K^+ \pi^-$ events within a $10$ MeV mass window of the $\phi$
meson peak in $K^- K^+$ invariant mass.
The inclusion of charge conjugate modes is implied
throughout this article unless otherwise stated.

We identify a ST in the $\psi(3770)$ data sample using the
energy difference
$\Delta E = E_D - E_\text{beam}$
and the
beam-constrained mass difference
$\Delta M_\text{bc} = [E^2_\text{beam} - {\bf p}_D^2]^{1/2} - m_D$,
where
$E_D$ is the energy of the tag,
$E_\text{beam}$ is the beam energy,
${\bf p}_D$ is the three momentum of the tag,
and
$m_D$ is the nominal mass~\cite{pdg2008}
of the neutral or charged charm meson.
We require the $\bar{D}^0 \to K^+ \pi^-$ and  $D^- \to K^+ \pi^- \pi^-$
tags to have $\Delta M_\text{bc}$ within a $4$ MeV mass window around the
nominal $D$ mass.

For data collected at the center-of-mass energy of $4170$ MeV,
we identify a ST by using
the invariant mass of the tag $M(D_s)$
and recoil mass against the tag $M_\text{recoil}(D_s)$.
The recoil mass is defined as
$M_\text{recoil}(D_s) =
[ (E_{ee} - E_{D_s} )^2 - ({\bf p}_{ee} - {\bf p}_{D_s})^2 ]^{1/2}
$,
where $(E_{ee}, {\bf p}_{ee})$ is the net four-momentum of the $e^+ e^-$ beam
taking the finite beam crossing angle into account,
and
$(E_{D_s},{\bf p}_{D_s})$ is the four-momentum of the tag,
with $E_{D_s}$ computed from ${\bf p}_{D_s}$ and
the nominal mass~\cite{pdg2008} 
of the $D_s$ meson.
We require the recoil mass to be
within $55$ MeV of the $D^\ast_s$ mass~\cite{pdg2008}.
This loose window allows both
primary and secondary
(from $D^{\ast -}_s \to D^-_s \gamma$ or $D^{\ast -}_s \to D^-_s \pi^0$)
$D_s$ tags to be selected.
We veto tag candidates with track momenta below $100$ MeV to
reduce the background from $D \bar{D}^{\ast}$ decays
(through $D^{\ast} \to \pi D$).

\begin{figure*}
\centering
\includegraphics[width=\textwidth]{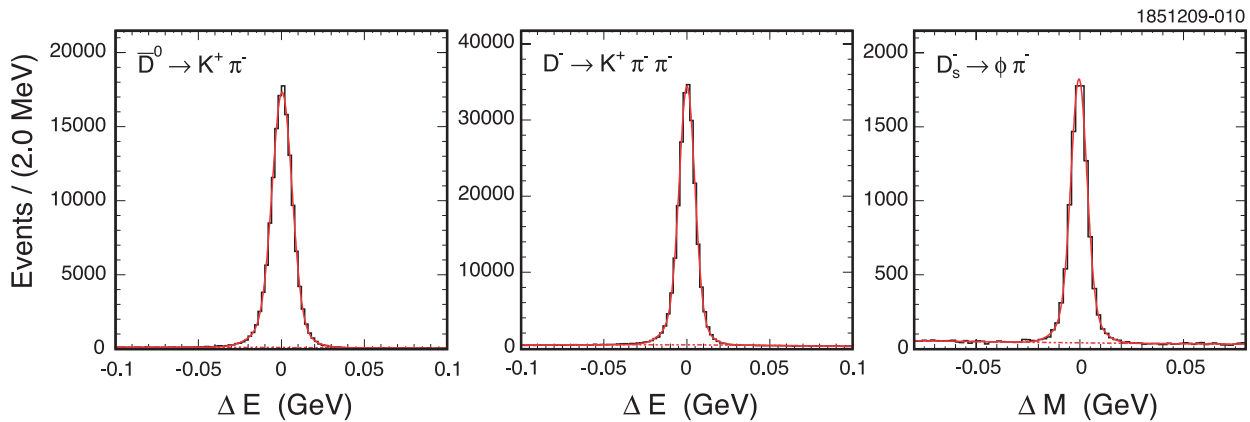}
\caption{\label{fig:st}Tag $\Delta E$ and $\Delta M$
distributions in data (histograms)
with fits (solid curves)
and
background contributions (dashed lines).
}
\end{figure*}

The $\Delta E$ and $\Delta M$ distributions
obtained from data are shown in
Fig.~\ref{fig:st}.
To estimate the backgrounds from the wrong tag combinations,
we use the sidebands of the $\Delta E$ distribution
or the tag mass difference $\Delta M = M(D_s) - m_{D_s}$ distribution,
where
$m_{D_s}$ is the nominal mass~\cite{pdg2008} of the $D_s$ meson.
We define the
signal and sideband regions in Table~\ref{table:signal-sideband-region}.
We fit the distributions to
a sum of a double-Gaussian function (for signal)
and a second order Chebyshev polynomial function (for background)
to determine the tag sideband scaling factor $s_\text{tag}$,
which is the ratio of areas in the signal and sideband regions
described by the background polynomial function.
Obtained ST yields and tag sideband scaling factors are listed in
Table~\ref{table:data-single}.

\begin{table}
\centering
\caption{\label{table:signal-sideband-region}
Signal and sideband regions of $\Delta E$ and $\Delta M$
for each tag mode.}
\begin{ruledtabular}
\begin{tabular}{lcc}
Tag mode&
Signal (MeV)&
Sideband (MeV)
\\
\hline
$\bar{D}^0 \to K^+ \pi^-$&
$-30 \le \Delta E < +30$ &
$-80 \le \Delta E < -50$ \\
 &
 &
$+50 \le \Delta E < +80$ \\
$D^- \to K^+ \pi^- \pi^-$&
$-25 \le \Delta E < +25$ &
$-65 \le \Delta E < -40$ \\
 &
 &
$+40 \le \Delta E < +65$ \\
$D^-_s \to \phi \pi^-$&
$-20 \le \Delta M < +20$ &
$-55 \le \Delta M < -35$ \\
 &
 &
$+35 \le \Delta M < +55$ \\
\end{tabular}
\end{ruledtabular}
\end{table}

\def\fitdesz{2.36in}
\def\fitdmsz{2.36in}

\begin{table}
\centering
\caption{\label{table:data-single}ST yields and statistical uncertainties
in data,
where $n^{S}_\text{ST}$ is the yield in the tag signal region,
$n^{B}_\text{ST}$ is the yield in the tag sideband region,
$s_\text{tag}$ is the tag sideband scaling factor obtained from a fit to tag
$\Delta E$ (or $\Delta M$) distribution,
and $n_\text{ST}$ is the scaled sideband subtracted ST yield.
}
\begin{ruledtabular}
\begin{tabular}{l rrrr}
Tag mode & $n^{S}_\text{ST}$& $n^{B}_\text{ST}$& $s_\text{tag}$& $n_\text{ST}$\\
\hline
$\bar{D}^{0} \to K^{+} \pi^{-}$ &
$144260$& $2258$& $1.067$& 
$141851 \pm 383$
\\
$D^{-} \to K^{+} \pi^{-} \pi^{-}$ &
$231429$& $7748$& $1.104$& 
$222872 \pm 490$
\\
$D_s^- \rightarrow \phi \pi^-$ &
$10453$& $807$& $0.979$& 
$9663 \pm 106$
\\
\end{tabular}
\end{ruledtabular}
\end{table}

\subsection{\label{sec:event_selectuon::signal_selection}Signal Selection}

We form DT candidates from ST candidates by adding a recoiling charged
track that is consistent with coming from the nominal interaction point.
Specifically,
the recoiling track's point of closest approach to the origin must be within
$5$ cm of the interaction point along the beam line and within $5$ mm of the
interaction point in the plane transverse to the beam line.
We require the momentum of the track to be
$p \ge 200$ MeV and the angle with respect to the beam to be
$|\cos{\theta}| < 0.80$ so that all charged-particle identification
(PID) information ($dE/dx$, RICH, and $E/p$) is available.
The signal track in the DT candidate
is also required to be identified as an electron, a charged pion,
or a charged kaon,
for further analysis.
This is discussed in the next section.

%
%
\section{\label{sec:analysis}Analysis}

%
%
The $D$ (or $D_s$) semileptonic inclusive spectrum (or differential
decay rate) can be expressed as
\begin{equation}
  \frac{d \mathcal{B}_{\text{SL}}}{d p}
	 =  \frac{1}{n_D} \frac{\Delta n_e}{\Delta p}
	 =  \frac{1}{n_\text{ST}}
	\frac{\Delta n_\text{DT}/\epsilon_\text{SL}}{\Delta p},
\label{eq:decay_rate}
\end{equation}
where
$n_D$ is 
the number of $D$ mesons produced,
$n_e$ is the number of produced primary electrons in bins of momentum $p$,
$n_\text{ST}$ is the number of ST,
$\Delta n_\text{DT}$ is the electron candidate yield in bins of momentum,
and $\epsilon_\text{SL}$ is the (momentum-dependent) electron detection
efficiency.
The $D$ semileptonic branching fraction can be obtained by integrating the
differential spectrum and correcting for the $200$ MeV momentum cutoff by
extrapolating the spectrum below the cutoff.
If we had a perfect MC modeling of the semileptonic decays,
a simple momentum bin-by-bin correction factor could be used for
$\epsilon_\text{SL}$.
Instead, we use a more general unfolding~\cite{Hocker:1995kb}
approach to minimize MC model dependence.

The observed laboratory momentum spectrum
$y(b, i_\text{track})$
of a particle identified as type $b$ ($= e$, $\pi$, or $K$)
in bins of measured track momentum bin $i_\text{track}$
can be modeled as a folded distribution.
It is related to the true laboratory momentum $n(a,j)$  via
detector-response matrices that account for resolution and efficiency:
\begin{widetext}
\begin{equation}
y(b, i_\text{track})
=
\sum_{a}
        A_\text{PID} (b| a, i_\text{track})
\sum_{j}
        A_\text{track} (i_\text{track} | a,j)
	n(a,j), \label{eq:model}
\end{equation}
\end{widetext}
where
$a$ ($= e$, $\mu$,  $\pi$, or $K$) is the true particle species index,
$n(a,j)$ is the true laboratory momentum spectrum in bins of
true laboratory momentum bin index $j$ of a particle type $a$,
$A_\text{track} (i_\text{track} | a,j)$
is the tracking efficiency matrix, which
describes the probability of a particle of type $a$
with momentum in bin $j$ to be
reconstructed in track momentum bin $i_\text{track}$,
and
$A_\text{PID} (b| a, i_\text{track})$
is the PID efficiency matrix, which describes the probability of a particle
of type $a$
with measured momentum
in bin $i_\text{track}$ to be identified as PID type $b$.
We unfold~\cite{Hocker:1995kb} Eq.~(\ref{eq:model}) to obtain
the true momentum spectrum
\begin{widetext}
\begin{eqnarray}
n (a=e,j)
=
\sum_{i_\text{track}}
        A^{-1}_\text{track} (i_\text{track} | a=e,j)
\left[
\sum_{b}
        A^{-1}_\text{PID} (b| a, i_\text{track})
	y(b, i_\text{track})
\right]_{a=e}, \label{eq:solution_lab}
\end{eqnarray}
\end{widetext}
where
the $A^{-1}$'s are the unfolded
inverses of each efficiency matrix.
Because we are interested in the primary electron laboratory momentum
spectrum (to obtain the branching fraction)
we use the electron solution after PID unfolding ($a = e$).

In addition to finite resolution and efficiency,
modeled by detector-response matrices, we have to
consider possible backgrounds in our observed spectrum.
We remove combinatorial wrong-tag background contribution
by $\Delta E$ (or $\Delta M$) sideband subtraction.
Charge symmetric nonprimary true electron backgrounds (from
$\gamma$ conversion and $\pi^0$ Dalitz decay)
are subtracted by using the wrong-sign (WS, opposite to the expected primary
electron charge) electron sample.
In the following subsections,
we break the analysis described above into discrete steps.

\subsection{\label{sec:yield}PID Yield}

From a set of signal candidate tracks,
we measure the PID yield
%
$y(b, i)$
in bins of PID type $b$,
track momentum bin $i_\text{track}$,
$\Delta E$ (or $\Delta M$) signal and sideband regions $i_\text{SB}$,
and right-sign (RS) or wrong-sign (WS) bin $i_\text{RW}$
depending on the charge of the track and the flavor of the tag,
where
$i$ is a collective index for $(i_\text{track}, i_\text{SB}, i_\text{RW})$.
The charge of the daughter kaon defines the flavor of the
$\bar{D}^0 \to K^+ \pi^-$ tag,
and the charge of the tag defines
$D^- \to K^+ \pi^- \pi^-$ and $D^-_s \to \phi \pi^-$ tags.
The RS track is defined to be the track with the same charge
as the tagged $\bar{D}^0$ daughter kaon or to be the opposite charge of the
charged tags, and the WS track is defined the other way around.

\subsection{\label{sec:Apid}PID Unfolding}

\begin{figure*}
\centering
\includegraphics[width=\textwidth]{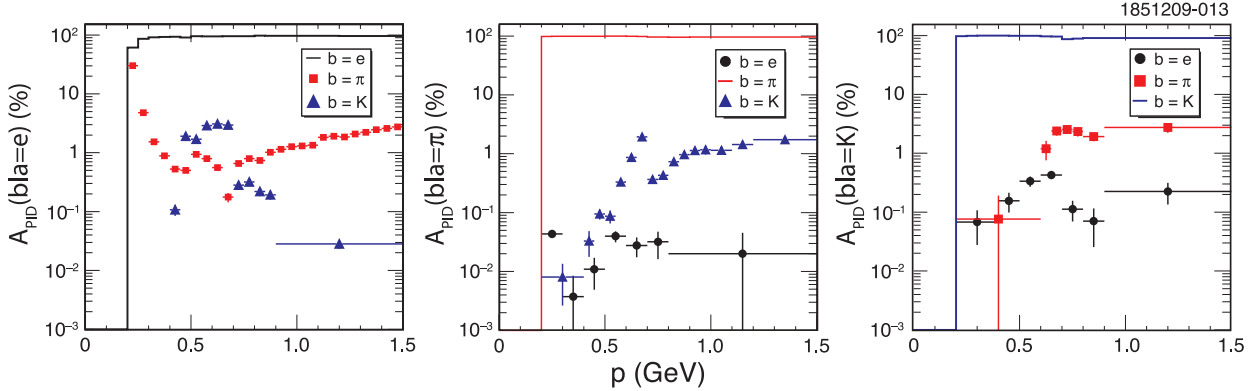}
\caption{\label{fig:apid-data}
(color online).
The components of the PID efficiency matrix
$A_\text{PID} (b|a)$ obtained from data.
The matrix
describes the probability of a particle of type $a$ to be identified
as a PID type $b$.
We measured the PID matrix in momentum
intervals of $50$ MeV (some bins are wider due to low statistics)
above the PID momentum cutoff $200$ MeV.
The cases with $a \neq b$, conventionally called the fake rate
or mis-PID probability, are shown in
points with statistical uncertainties.
The cases with $a = b$, conventionally called the efficiency,
are shown as solid lines.
The discontinuities at momentum $700$ MeV
in fake rates and efficiencies are due to the fact that
the RICH information is used for pion and kaon identifications
only above $700$ MeV.
}
\end{figure*}

We correct for PID
efficiency and mis-PID crossfeed backgrounds using
\begin{equation}
y (a, i)
=
        A^{-1}_\text{PID} (b| a, i)
	y(b, i), \label{eq:pid_problem}
\end{equation}
where
$i$ is a collective index for $(i_\text{track}, i_\text{SB}, i_\text{RW})$.
The PID matrix
$A_\text{PID}(b|a)$ used in the unfolding is shown in Fig.~\ref{fig:apid-data}.
PID matrix elements associated with
the charged pion are obtained from $K^0_S \to \pi^+ \pi^-$ events,
the charged kaon elements are obtained from $D^+ \to K^- \pi^+ \pi^+$ events,
and
the electron elements are obtained from
radiative Bhabha events ($e^+ e^- \gamma$)
embedded in hadronic events.
Here we treat muons as pions because muons in the momentum range
in which we are interested behave almost
the same as charged pions in the CLEO-c detector.
The effect of this approximation is negligible
on our branching fraction measurement because the probability of
pions (and muons) to be misidentified as electrons is very small,
as shown in Fig.~\ref{fig:apid-data}.
After solving the PID problem, we take the electron solution ($a = e$)
for further analysis.

\def\apidsz{6cm}

\subsection{\label{sec:sideband_subtraction}Tag Sideband Subtraction}

To remove the wrong-tag combinatorial background,
we perform $\Delta E$ (or $\Delta M$) sideband subtractions
after PID unfolding.
After this process, we deal with real
electrons from $D$ (or $D_s$) meson decay.

\subsection{\label{sec:wrongsign_subtraction}Wrong-Sign Electron Subtraction}

Charge symmetric secondary electrons are removed by subtracting
the WS (secondary) electron yield from the RS (primary plus
secondary) electron yield.
After this process, we end up with primary
electrons from  $D$ (or $D_s$) meson decay.

\subsection{\label{sec:Atrack}Tracking Efficiency, $A_\text{track}$}

We obtain the tracking efficiency matrix $A_\text{track} (i_\text{track}|j)$
from MC simulation. This includes
track finding efficiency and resolution effects.

\subsection{\label{sec:tag_bias}Tag Bias Correction}

The signal semileptonic efficiency $\epsilon_\text{SL}$
requires a possible tag bias correction
which would be introduced if the ST efficiency in the signal DT events
is different from that when the other recoiling system is
a generic $D$-meson (or $D_s$-meson) decay.  The effect of the tag bias
can be express in terms of a ST efficiency ratio
\begin{equation}
\epsilon_\text{SL}
=
\frac{\epsilon_\text{DT}}{\epsilon_\text{ST}}
=
\frac{\epsilon_\text{DT}}{\epsilon^\prime_\text{ST}}
\frac{\epsilon^\prime_\text{ST}}{\epsilon_\text{ST}}
=
\frac{\epsilon_{e} \epsilon^\prime_\text{ST}}{\epsilon^\prime_\text{ST}}
\frac{\epsilon^\prime_\text{ST}}{\epsilon_\text{ST}}
=
\epsilon_{e} b_\text{tag}
,
\end{equation}
where
$\epsilon_\text{DT}$ is the DT efficiency,
$\epsilon_\text{ST}$ is the ST efficiency against generic decays in the
recoiling system,
$\epsilon^\prime_\text{ST}$ is the ST efficiency when the recoiling
system is the signal semileptonic decays,
$\epsilon_{e}$ is the signal electron detection efficiency
given the tag in the other side is found,
and $b_\text{tag}$ is a measure of tag bias in the efficiency
Thus, $b_\text{tag} = \epsilon^\prime_\text{ST} / \epsilon_\text{ST}$
and $\epsilon_{e} = \epsilon_\text{DT} / \epsilon^\prime_\text{ST}$.
We expect this effect to be small due to chosen clean tag modes
and low event multiplicity.
We estimate tag biases in MC simulation:
$b_\text{tag} (D^0 \to e^+ X) = 0.9965 \pm 0.0017$,
$b_\text{tag} (D^+ \to e^+ X) = 1.0017 \pm 0.0021$,
and
$b_\text{tag} (D^+_s \to e^+ X) = 1.0069 \pm 0.0021$,
where uncertainties are due to MC statistics.

\subsection{\label{sec:dcsd-correction}Doubly Cabibbo-suppressed Decay Correction}

Because of the doubly Cabibbo-suppressed decay (DCSD)
and quantum correlation~\cite{Asner:2005wf,Rosner:2008fq}
in coherent $D^0 \bar{D}^0$  production
at the $\psi(3770)$ resonance energy, we need a correction for the
observed semileptonic branching fraction
using the $\bar{D}^0 \to K^- \pi^+$ tag mode.
The observed branching fraction $\mathcal{B}_\text{obs}$ requires
a correction~\cite{Asner:2005wf,Rosner:2008fq}
\begin{equation}
\mathcal{B}(D^0 \to X e^+ \nu_e) =
\frac{1+R_\text{WS}}{1-r^2}
\mathcal{B}_\text{obs}(D^0 \to X e^+ \nu_e).
\end{equation}
Here
$r^2 = |\langle K^+\pi^-| D^0\rangle / \langle K^+ \pi^-| \bar{D}^0\rangle|^2$
is the ratio of the DCSD
rate to the Cabibbo-favored decay rate,
and
$R_\text{WS}= \Gamma(D^0 \to K^+ \pi^-)/\Gamma(\bar{D}^0 \to K^+ \pi^-)$
is the ratio of the time-integrated DCSD rate to the
Cabibbo-favored decay rate.
Using the world average~\cite{pdg2008} values of these
we need a correction factor
$(1+R_\text{WS})/(1-r^2)
= [1 + (3.80 \pm 0.05)\times 10^{-3}]/[1 - (3.35 \pm 0.09)\times 10^{-3}]
= 1.0072 \pm 0.0001
$.

%
%
\section{\label{sec:results}Results}

\begin{table*}
\centering
\caption{\label{table:yield}
Summary of DT yields, statistical uncertainties,
and correction procedure explained in
Sec.~\ref{sec:analysis}.
PID yields (Sec.~\ref{sec:yield}) for electron candidates ($b = e$) are
shown in the first group
for tag signal region ($S$), tag sideband region ($B$),
right-sign ($R$), and wrong-sign ($W$) bins,
where the yields in the sideband region are scaled by the tag sideband scaling
factor (Table~\ref{table:data-single}) for each tag mode.
PID unfolded (Sec.~\ref{sec:Apid}) electron yields ($a = e$)
are shown in the second group.
Tag sideband subtracted (Sec.~\ref{sec:sideband_subtraction})
electron yields are shown in the third group,
followed by the
wrong-sign subtracted yield (Sec.~\ref{sec:wrongsign_subtraction}),
tracking efficiency-corrected yield (Sec.~\ref{sec:Atrack}),
and remaining tag bias (Sec.~\ref{sec:tag_bias}) or
DCSD (Sec.~\ref{sec:dcsd-correction}) corrected yield.
}
\begin{ruledtabular}
\begin{tabular}{l r r r}
 & $D^{0}$& $D^{+}$& $D^{+}_{s}$\\
\hline
PID yield, electron candidates& & & \\
\hspace{0.25in} $y(b=e,S,R)$&$6618.0 \pm 81.4$& $24834.0 \pm 157.6$& $553.0 \pm 23.5$\\
\hspace{0.25in} $y(b=e,B,R)$&$41.6 \pm 6.7$& $332.4 \pm 19.2$& $24.5 \pm 4.9$\\
\hspace{0.25in} $y(b=e,S,W)$&$653.0 \pm 25.6$& $711.0 \pm 26.7$& $50.0 \pm 7.1$\\
\hspace{0.25in} $y(b=e,B,W)$&$19.2 \pm 4.5$& $55.2 \pm 7.8$& $9.8 \pm 3.1$\\
PID unfolded yield, electrons& & & \\
\hspace{0.25in} $y(a=e,S,R)$&$7292.4 \pm 90.7$& $27304.5 \pm 174.8$& $608.9 \pm 26.4$\\
\hspace{0.25in} $y(a=e,B,R)$&$47.1 \pm 7.7$& $370.4 \pm 21.7$& $27.7 \pm 5.6$\\
\hspace{0.25in} $y(a=e,S,W)$&$682.4 \pm 31.4$& $812.8 \pm 33.8$& $56.7 \pm 8.6$\\
\hspace{0.25in} $y(a=e,B,W)$&$21.3 \pm 5.3$& $65.2 \pm 9.8$& $11.7 \pm 3.4$\\
Tag sideband subtracted electrons& & & \\
\hspace{0.25in} $y(a=e,R)$&$7245.3 \pm 91.0$& $26934.1 \pm 176.2$& $581.2 \pm 27.0$\\
\hspace{0.25in} $y(a=e,W)$&$661.1 \pm 31.9$& $747.6 \pm 35.2$& $44.9 \pm 9.2$\\
Wrong-sign subtracted electrons&$6584.2 \pm 96.4$& $26186.5 \pm 179.6$& $536.3 \pm 28.5$\\
Tracking efficiency-corrected electrons&$8361.0 \pm 123.0$& $33182.0 \pm 228.2$& $681.3 \pm 36.4$\\
Tag bias (and DCSD) corrected electrons&$8450.8 \pm 124.3$& $33125.6 \pm 227.9$& $676.6 \pm 36.2$\\
\end{tabular}
\end{ruledtabular}
\end{table*}


\def\specsz{2.36in}
\begin{figure*}
\centering
\includegraphics[width=\textwidth]{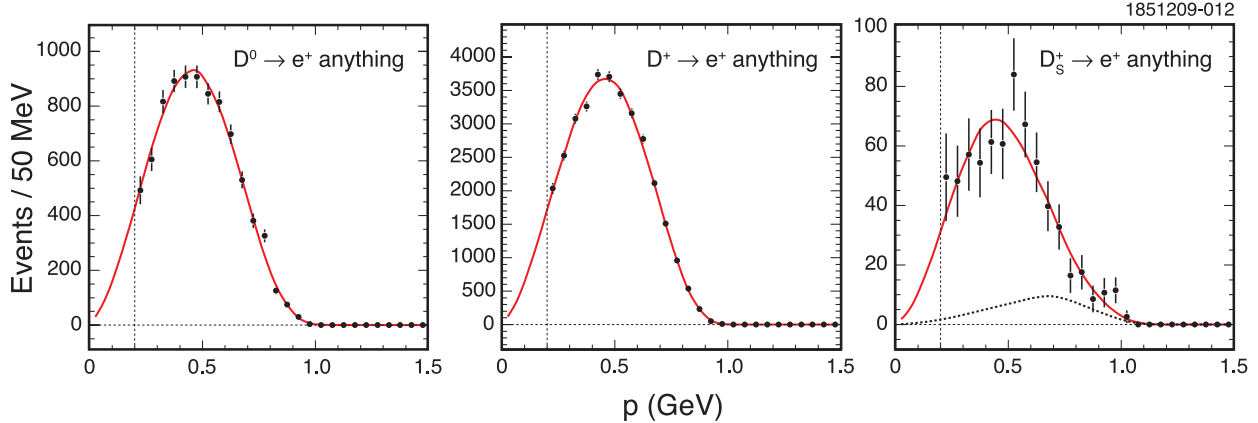}
\caption{\label{fig:spectra}
Inclusive laboratory frame
electron spectra obtained from data, shown as points with statistical
uncertainties.
The vertical dashed lines indicate the PID momentum cutoff at $200$ MeV.
Extrapolated spectra are shown as solid curves.
The dashed curve in the $D^+_s$ spectrum plot is the
expected contribution from
$\tau^+ \to e^+ \nu_e \bar{\nu}_\tau$ 
from
leptonic $D^+_s \to \tau^+ \nu_\tau$ decay.
}
\end{figure*}
\begin{table*}
\caption{\label{table:br-result}
Summary of semileptonic branching fractions.
Here
$\mathcal{B}_\text{trunc}$ is the partial branching fraction above $200$ MeV,
$\mathcal{B}(e^+ X)$ is the extrapolated full branching fraction,
and $\mathcal{B}(X e^+ \nu_e)$ is the semileptonic branching fraction
after $\tau \to e$ correction (for $D^+$ and $D^+_s$).
First uncertainties are statistical and the second are systematic
due to uncertainties in
%
$\mathcal{B}(D^+ \to \tau^+ \nu_\mu)$~\cite{:2008sq},
$\mathcal{B}(D^+_s \to \tau^+ \nu_\tau)$~\cite{Alexander:2009ux,Onyisi:2009th},
and
$\mathcal{B}(\tau^+ \to e^+ \nu_e \bar{\nu}_\tau)$~\cite{pdg2008}.
}
\centering
\begin{ruledtabular}
\begin{tabular}{l rrr}
Tag mode& $B_\text{trunc}(e^+ X)$ (\%)& $B(e^+ X)$ (\%)& $B(X e^+ \nu_e)$ (\%)\\
\hline
$\bar{D}^{0} \to K^{+} \pi^{-}$&
$5.958 \pm 0.084$& $6.460 \pm 0.091$& $6.460 \pm 0.091$
\\
$D^{-} \to K^{+} \pi^{-} \pi^{-}$&
$14.863 \pm 0.092$& $16.147 \pm 0.100$& $16.129 \pm 0.100 \pm 0.000$
\\
\vspace{-1em}
\\
$D_s^- \rightarrow \phi \pi^-$&
$7.002 \pm 0.361$& $7.525 \pm 0.387$& $6.522 \pm 0.387 \pm 0.079$
\\
\end{tabular}
\end{ruledtabular}
\end{table*}

The final electron candidate yields are summarized in Table~\ref{table:yield}
and
efficiency-corrected laboratory momentum spectra are shown
in Fig.~\ref{fig:spectra}.
Also shown in Fig.~\ref{fig:spectra}
are the spectrum extrapolations below the PID
momentum cutoff (200 MeV).
The curves shown are obtained with a fit using
the sum of measurements of exclusive channels together with form-factor
models and adding higher-resonance and nonresonant channels to match the
sum of the exclusive channels with our measured branching fraction.
Further details of the
extrapolation procedure are available in the Appendix.
From the fit results, we obtain
fractions below the momentum cutoff of
$7.8\%$ for $D^0$,
$8.0\%$ for $D^+$,
and
$7.0\%$ for $D^+_s$.

At this point, we also consider the
secondary electrons from leptonic decays of $D^+ \to \tau^+ \nu_\tau$
and $D^+_s \to \tau^+ \nu_\tau$ as they produce electrons through
$\tau^+ \to e^+ \nu_e \bar{\nu}_\tau$ decay.
This source of secondary electrons is expected to be large
in $D^+_s$, so we have included the expected spectrum component in the
extrapolation.
The expected branching fractions
of these secondary electrons from the leptonic decays of $D^+$ and $D^+_s$
are subtracted from the fully inclusive branching fraction results
to obtain
inclusive semileptonic decay branching fractions.
The branching fraction for $D^+_s \to \tau^+ \nu_\tau$ decay
is taken from Refs.~\cite{Alexander:2009ux,Onyisi:2009th},
$\mathcal{B}(D^+_s \to \tau^+ \nu_\tau) = (5.62 \pm 0.41 \pm 0.16)\%$.
The size of the expected secondary electron contribution
from the unobserved leptonic decay $D^+ \to \tau^+ \nu_\tau$
is based on the known branching fraction of
$D^+ \to \mu^+ \nu_\mu$ decay~\cite{:2008sq} scaled by the
standard model decay rate ratio~\cite{pdg2008}
$\Gamma(D^+ \to \tau^+ \nu_\tau) / \Gamma(D^+ \to \mu^+ \nu_\mu) = 2.67$.
We take the uncertainty in the $\tau \to e$ correction as a part of our
systematic uncertainty.
Branching fraction results are summarized in Table~\ref{table:br-result}
with all above-mentioned efficiency and cutoff corrections.

The laboratory frame electron momentum spectra
shown in Fig.~\ref{fig:spectra} are given in tabular form in
Table~\ref{table:spectra}.

\begin{table*}
\centering
\caption{\label{table:spectra}
Inclusive semileptonic electron partial branching fractions
of $D^0$, $D^+$, and $D^+_s$
in the laboratory frame.  For $D^+$ and $D^+_s$, we have subtracted
expected contributions from leptonic decays $\tau^+ \nu_\tau$
(followed by $\tau^+ \to e^+ \nu_e \bar{\nu_\tau}$).
Systematic uncertainties in total branching fractions are added to the
statistical uncertainties.
In comparing theoretical predictions with these measurements,
one must smear the theoretical predictions by boosting from
the $D$ (or $D_s$) rest frame to the laboratory frame.  For $D_s$,
$51\%$ of the electrons are from secondary $D_s$ from $D^\ast_s$,
and $49\%$ are from primary $D_s$.}
\begin{ruledtabular}
\begin{tabular}{l rrr}
$p$ (GeV)& $\Delta B(D^0 \to X e^+ \nu_e)$ (\%)& $\Delta B(D^+ \to X e^+ \nu_e)$ (\%)& $\Delta B(D^+_s \to X e^+ \nu_e)$ (\%)\\
\hline
$0.200$--$0.250$& $0.347 \pm 0.036$ &$0.912 \pm 0.040$ &$0.491 \pm 0.152$
\\
$0.250$--$0.300$& $0.426 \pm 0.030$ &$1.133 \pm 0.038$ &$0.470 \pm 0.124$
\\
$0.300$--$0.350$& $0.576 \pm 0.031$ &$1.379 \pm 0.041$ &$0.554 \pm 0.126$
\\
$0.350$--$0.400$& $0.629 \pm 0.030$ &$1.462 \pm 0.043$ &$0.515 \pm 0.120$
\\
$0.400$--$0.450$& $0.640 \pm 0.031$ &$1.675 \pm 0.047$ &$0.578 \pm 0.112$
\\
$0.450$--$0.500$& $0.640 \pm 0.031$ &$1.661 \pm 0.046$ &$0.562 \pm 0.123$
\\
$0.500$--$0.550$& $0.596 \pm 0.029$ &$1.546 \pm 0.044$ &$0.794 \pm 0.127$
\\
$0.550$--$0.600$& $0.575 \pm 0.029$ &$1.415 \pm 0.041$ &$0.611 \pm 0.115$
\\
$0.600$--$0.650$& $0.492 \pm 0.026$ &$1.243 \pm 0.038$ &$0.471 \pm 0.104$
\\
$0.650$--$0.700$& $0.374 \pm 0.023$ &$0.946 \pm 0.032$ &$0.314 \pm 0.087$
\\
$0.700$--$0.750$& $0.269 \pm 0.019$ &$0.674 \pm 0.026$ &$0.246 \pm 0.079$
\\
$0.750$--$0.800$& $0.230 \pm 0.017$ &$0.429 \pm 0.019$ &$0.089 \pm 0.060$
\\
$0.800$--$0.850$& $0.089 \pm 0.011$ &$0.240 \pm 0.014$ &$0.115 \pm 0.060$
\\
$0.850$--$0.900$& $0.053 \pm 0.008$ &$0.103 \pm 0.009$ &$0.037 \pm 0.046$
\\
$0.900$--$0.950$& $0.021 \pm 0.005$ &$0.022 \pm 0.004$ &$0.074 \pm 0.051$
\\
$0.950$--$1.000$& $0.002 \pm 0.002$ &$0.004 \pm 0.002$ &$0.096 \pm 0.045$
\\
$1.000$--$1.050$& $\cdot\cdot\cdot$ &  $\cdot\cdot\cdot$& $0.015 \pm 0.022$
\\
\end{tabular}
\end{ruledtabular}
\end{table*}

%
%
\section{\label{sec:systematic_uncertainty}Systematic Uncertainties}

Possible sources of systematic uncertainties and
their effects on the branching fraction measurements
are summarized in
Table~\ref{table:systematic-uncertainty}.

\begin{table}
\centering
\caption{\label{table:systematic-uncertainty}
Summary of sources of systematic uncertainty and their effects on
the semileptonic branching fraction measurements.}
\begin{ruledtabular}
\begin{tabular}{lccc}
Source&
  $D^0$ (\%)&
  $D^+$ (\%)&
  $D^+_s$ (\%)
\\
\hline
Number of tags&
$0.5$ & $0.7$& $0.9$\\
%
Tracking&
$0.3$ & $0.3$& $0.3$\\
PID&
$0.8$ & $0.5$& $0.6$\\
%
FSR&
$0.5$& $0.5$& $0.5$\\
Tag bias&
$0.2$ & $0.2$& $0.3$\\
DCSD&
$0.0$ & $\cdot\cdot\cdot$ & $\cdot\cdot\cdot$\\
$\tau \to e$&
$\cdot\cdot\cdot$ & $0.0$ & $1.2$\\
Extrapolation&
$1.3$ & $1.4$& $1.5$\\
%
\hline
Total&
$1.7$ & $1.8$& $2.3$\\
\end{tabular}
\end{ruledtabular}
\end{table}

%
%
The ST yields are obtained from a tag ($\Delta E$ or $\Delta M$)
sideband subtraction method.  Because of the chosen clean tag modes,
there is very little combinatorial background under the signal peak,
as shown in Fig.~\ref{fig:st}.
Systematic uncertainties in the numbers of tags
are studied by using alternative signal and background
functions, and comparing
the known input number of ST in a MC simulation test
to the output with the same procedure.
By adding all of the resulting variations in quadrature, we obtain
$0.5\%$ (in $D^0$),
$0.7\%$ (in $D^+$),
and $0.9\%$ (in $D^+_s$)
uncertainties in the estimation of the
number of ST.

%
%
The systematic uncertainty of $0.3$\% in tracking efficiency
was estimated~\cite{:2007zt}
in a detailed MC and data efficiency comparison using
$\psi(3770) \to D\bar{D}$ events with the cases when both $D$ and
$\bar{D}$ mesons can be fully reconstructed.

%
%
%
Uncertainties in FSR and bremsstrahlung effects
on $D$ semileptonic decay branching
fraction measurements were studied in
our previous measurement~\cite{Adam:2006nu}
and in high statistics exclusive
$D$ semileptonic decay modes~\cite{Besson:2009uv}.
They are found to be well simulated in our MC program.
We have assessed the uncertainty in FSR by redoing the
analysis using alternative
signal efficiency and input spectra
with FSR turned off in the MC simulation.
Including the uncertainty in bremsstrahlung simulation~\cite{Besson:2009uv},
we assign $0.5\%$ uncertainty due to FSR and bremsstrahlung effects
on our branching fraction measurements.

%
%
Uncertainties in electron identification
for semileptonic decays are assessed
by comparing the efficiency measured using a radiative Bhabha sample
embedded in hadronic events to those in various MC simulated event samples.
We assign systematic uncertainties due to electron identification
as
$0.7\%$ for $D^0 \to X e^+ \nu_e$,
$0.5\%$ for $D^+ \to X e^+ \nu_e$,
and
$0.6\%$ for $D^+_s \to X e^+ \nu_e$.
For other PID efficiencies,
we have varied their values within measured uncertainties
and observe the effect on our measured branching fraction.
By adding all electron identification and other PID uncertainties in quadrature we assign
uncertainties in PID as
$0.8\%$ for $D^0 \to X e^+ \nu_e$,
$0.5\%$ for $D^+ \to X e^+ \nu_e$,
and
$0.6\%$ for $D^+_s \to X e^+ \nu_e$.

%
%
Tag bias corrections are estimated from MC simulation.
We take the uncertainty in the MC statistics and
a quarter of the size of the tag bias
as the uncertainty in the correction.

%
%
For the $D^+_s$ and $D^+$ inclusive electron spectra, we subtract
the contribution from $\tau^+ \to e^+ \nu_e \bar{\nu}_\tau$
as estimated in Table~\ref{table:data-dpsl}
(and Table~\ref{table:data-dssl})
to obtain the inclusive semileptonic branching fraction.
We take the uncertainty from $\mathcal{B}(D^+_s \to \tau^+ \nu_\tau)$
[and $\mathcal{B}(D^+ \to \tau^+ \nu_\tau)$]
listed in the table for the uncertainty on the $\tau \to e$ contribution
correction. The uncertainty in $D^+ \to X e^+ \nu_e$ is negligible,
and we assign an uncertainty of $1.2$\% in $D^+_s \to X e^+ \nu_e$.

%
%
To estimate systematic uncertainties in the extrapolation procedure,
we fix all parameters to the reference values listed in
Table~\ref{table:data-d0sl}, Table~\ref{table:data-dpsl},
and Table~\ref{table:data-dssl}.
Then we vary each semileptonic decay component one-by-one
within the allowed range of uncertainties, listed in the table,
and reevaluate the fraction below the momentum cutoff and the
effect on the resulting branching fraction.
For the unobserved decay components, we vary
$100\%$ of the size of the predicted branching fraction
to assess the uncertainty.
We also use alternative form-factor models, by changing
models component-by-component from the reference models in the tables,
when we perform an extrapolation fit as described in the Appendix,
to assess the additional uncertainty in the extrapolation.
By adding all effects in quadrature,
we assign 
$1.3\%$ for $D^0 \to X e^+ \nu_e$,
$1.4\%$ for $D^+ \to X e^+ \nu_e$,
and
$1.5\%$ for $D^+_s \to X e^+ \nu_e$
as uncertainties in the extrapolation procedure.

%
%
\section{\label{sec:conclusion}Summary}


Using the full sample of open-charm data collected by the CLEO-c
detector, we obtain the charm and charmed-strange meson inclusive semileptonic
branching fractions:
\begin{eqnarray*}
\mathcal{B}(D^0 \to X e^+ \nu_e) & = & ( 6.46 \pm 0.09 \pm 0.11)\%,\\
\mathcal{B}(D^+ \to X e^+ \nu_e) & = & (16.13 \pm 0.10 \pm 0.29)\%,
\end{eqnarray*}
and
\begin{eqnarray*}
\mathcal{B}(D^+_s \to X e^+ \nu_e) & = & (6.52 \pm 0.39 \pm 0.15)\%,
\end{eqnarray*}
where the first uncertainties are statistical and the second are systematic.
Using known~\cite{pdg2008} lifetimes
$\tau_{D^0} = (410.1 \pm 1.5) \times 10^{-15}$ s,
$\tau_{D^+} = (1040 \pm 7) \times 10^{-15}$ s,
and
$\tau_{D^+_s} = (500 \pm 7) \times 10^{-15}$ s, we obtain the ratios of
semileptonic decay widths
\[
\frac{\Gamma (D^+ \to X e^+ \nu_e)}
     {\Gamma (D^0 \to X e^+ \nu_e)}
=
0.985 \pm 0.015 \pm 0.024
\]
and
\[
\frac{\Gamma (D^+_s \to X e^+ \nu_e)}
     {\Gamma (D^0 \to X e^+ \nu_e)}
=
0.828 \pm 0.051 \pm 0.025.
\]
In these ratios,
we assume the PID and tracking
uncertainties are fully correlated and all others are uncorrelated.
The former ratio shows that charged and neutral charm meson semileptonic
decay widths are consistent with isospin symmetry, as expected,
because the two mesons differ only in the isospin of the light quark.
On the other hand, the latter ratio shows that there is an indication of
difference between charm and charmed-strange meson semileptonic
decay widths.


%
\begin{acknowledgments}
We gratefully acknowledge the effort of the CESR staff
in providing us with excellent luminosity and running conditions.
D.~Cronin-Hennessy thanks the A.P.~Sloan Foundation.
This work was supported by
the National Science Foundation,
the U.S. Department of Energy,
the Natural Sciences and Engineering Research Council of Canada, and
the U.K. Science and Technology Facilities Council.
\end{acknowledgments}

\appendix*

\section{\label{app:extrapolation}Spectrum Extrapolation}

Charm and charmed-strange
exclusive semileptonic decay
components used to perform spectrum extrapolation
fits are summarized in
Tables \ref{table:data-d0sl},
\ref{table:data-dpsl},
and
\ref{table:data-dssl}.
Efficiency-corrected data points are fit to
a sum of exclusive semileptonic decay components to estimate
the unmeasured portion of the spectrum below the momentum cutoff at $200$ MeV
due to the electron identification.
Normalization of each component is allowed to float within the
uncertainty shown in the tables.

Higher-resonance and nonresonant decay components are used to
make the sum of exclusive branching fractions
match the inclusive branching fraction
in $D^0$ and $D^+$ extrapolations.  Higher-resonance decay branching fractions
are predicted by the ISGW2~\cite{Scora:1995ty} form-factor model and remaining
gaps are filled by nonresonant decays.
We assume the size
of the nonresonant component of $D \to \bar{K} \pi e^+ \nu_e$
to be about $5\%$~\cite{Link:2005ge,pdg2008}.
Uncertainties of unobserved higher-resonance channels are
assumed to be $\pm 100$\% of the predicted branching fractions.

The expected leptonic decay contributions due to the
$\tau^+ \to e^+ \nu_e \bar{\nu}_\tau$ decay in $D^+$ and $D^+_s$
are used to correct nonsemileptonic electrons in our measurements
as shown in Tables
\ref{table:data-dpsl}
and
\ref{table:data-dssl}.
For $D^+_s$ decays, this component is expected to be
large, and we include the leptonic decay component in the extrapolation fit.

\begingroup
\begin{table*}
\caption{\label{table:data-d0sl}
Summary of $D^0$ semileptonic decays used to perform
the spectrum extrapolation.
Assumed branching fractions are shown in the second column;
normalization of each component is allowed to float within the
given uncertainty.
Form-factor models used to describe the
shape of each spectrum are shown in the third column:
single-pole (SPOLE~\cite{Becirevic:1999kt}),
modified-pole (BK~\cite{Becirevic:1999kt}),
ISGW2~\cite{Scora:1995ty},
and phase space (PHSP).
Higher-resonance (and nonresonant) channels are used to match
the sum of exclusive semileptonic branching fractions to
the inclusive semileptonic branching fraction.
}
\begin{ruledtabular}
\begin{tabular}{llll}
Channel& $\mathcal{B}$ (\%)&
Form factor& Comment\\
\hline
$D^{0} \to K^{\ast -} e^{+} \nu_{e}$&
$2.16(17)$~\cite{Coan:2005iu}&
SPOLE &
$r_V = 1.62(8)$ and $r_2 = 0.83(5)$~\cite{pdg2008}
\\
$D^{0} \to K^{-} e^{+} \nu_{e}$&
$3.50(5)$~\cite{Besson:2009uv}&
BK &
%
$\alpha_\text{BK} = 0.30(3)$~\cite{Besson:2009uv}
\\
$D^{0} \to K^{-}_{1} e^{+} \nu_{e}$&
$0.11(11)$&
ISGW2 &
$\mathcal{B}$
from Ref.~\cite{Scora:1995ty} scaled by Ref.~\cite{Besson:2009uv}
\\
$D^{0} \to K^{\ast -}_{2} e^{+} \nu_{e}$&
$0.11(11)$&
ISGW2 &
$\mathcal{B}$
set to same as $D^{0} \to K^{-}_{1} e^{+} \nu_{e}$
\\
$D^{0} \to \bar{K} \pi e^{+} \nu_{e}$&
$0.12(3)$~\cite{Link:2005ge,pdg2008}&
PHSP &
Nonresonant
\\
$D^{0} \to \pi^{-} e^{+} \nu_{e}$&
$0.288(9)$~\cite{Besson:2009uv}&
BK &
%
$\alpha_\text{BK} = 0.21(7)$~\cite{Besson:2009uv}
\\
$D^{0} \to \rho^{-} e^{+} \nu_{e}$&
$0.16(2)$~\cite{ichep06:ygao}&
SPOLE &
$r_V = 1.4(3)$ and $r_2 = 0.6(2)$~\cite{ichep06:ygao}
\\
\end{tabular}
\end{ruledtabular}
\end{table*}
\endgroup

\begingroup
\begin{table*}
\caption{\label{table:data-dpsl}
Summary of $D^+$ semileptonic decays used to perform
the spectrum extrapolation.
Assumed branching fractions are shown in the second column;
normalization of each component is allowed to float within the
given uncertainty.
Form-factor (FF) models used to describe the
shape of each spectrum are shown in the third column:
single-pole (SPOLE~\cite{Becirevic:1999kt}),
modified-pole (BK~\cite{Becirevic:1999kt}),
ISGW2~\cite{Scora:1995ty},
and phase space (PHSP).
Higher-resonance (and nonresonant) channels are used to match
the sum of exclusive semileptonic branching fractions to the
inclusive semileptonic branching fraction.
The size of the expected secondary electron contribution
from the leptonic decay $D^+ \to \tau^+ \nu_\tau$ is shown in the last row
based on the known branching fraction of
$D^+ \to \mu^+ \nu_\mu$ decay~\cite{:2008sq} scaled by the
standard model decay rate ratio~\cite{pdg2008}
$\Gamma(D^+ \to \tau^+ \nu_\tau) / \Gamma(D^+ \to \mu^+ \nu_\mu) = 2.67$.
}
\begin{ruledtabular}
\begin{tabular}{llll}
Channel& $\mathcal{B}$ (\%)&
Form factor& Comment\\
\hline
$D^{+} \to \bar{K}^{\ast 0} e^{+} \nu_{e}$&
$5.56(35)$~\cite{Huang:2005iv}&
SPOLE &
$r_V = 1.62(8)$ and $r_2 = 0.83(5)$~\cite{pdg2008}
\\
$D^{+} \to \bar{K}^{0} e^{+} \nu_{e}$&
$8.83(22)$~\cite{Besson:2009uv}& 
BK &
%
$\alpha_\text{BK} = 0.30(2)$~\cite{pdg2008}
\\
$D^{+} \to \bar{K}^{0}_{1} e^{+} \nu_{e}$&
$0.29(29)$&
ISGW2 &
$\mathcal{B}$
from
Ref.~\cite{Scora:1995ty} scaled by Ref.~\cite{Besson:2009uv}
\\
$D^{+} \to \bar{K}^{\ast 0}_{2} e^{+} \nu_{e}$&
$0.29(29)$&
ISGW2 &
$\mathcal{B}$
set to same as $D^{+} \to \bar{K}^{0}_{1} e^{+} \nu_{e}$
\\
$D^{+} \to \bar{K} \pi e^{+} \nu_{e}$&
$0.32(8)$~\cite{Link:2005ge,pdg2008}&
PHSP &
Nonresonant
\\
$D^{+} \to \pi^{0} e^{+} \nu_{e}$&
$0.405(18)$~\cite{Besson:2009uv}&
BK &
%
$\alpha_\text{BK} = 0.21(7)$~\cite{Besson:2009uv}
\\
$D^{+} \to \eta e^{+} \nu_{e}$&
$0.13(2)$~\cite{Mitchell:2008kb}&
BK &
FF
set to same as
$D^{+} \to \pi^{0} e^{+} \nu_{e}$~\cite{Besson:2009uv}
\\
$D^{+} \to \eta^{\prime} e^{+} \nu_{e}$&
$0.02(2)$~\cite{Mitchell:2008kb,Scora:1995ty,Fajfer:2004mv}&
BK &
FF
set to same as
$D^{+} \to \pi^{0} e^{+} \nu_{e}$~\cite{Besson:2009uv}
\\
$D^{+} \to \rho^{0} e^{+} \nu_{e}$&
$0.23(2)$~\cite{ichep06:ygao}&
SPOLE &
$r_V = 1.4(3)$ and $r_2 = 0.6(2)$~\cite{ichep06:ygao}
\\
$D^{+} \to \omega e^{+} \nu_{e}$&
$0.15(3)$~\cite{ichep06:ygao}&
SPOLE &
FF
set to same as
$D^{+} \to \rho^{0} e^{+} \nu_{e}$~\cite{ichep06:ygao}
\\
\hline
%
$D^+ \to \tau^+ \nu_\tau$, $\tau^+ \to e^+ \nu_e \bar{\nu}_\tau$ &
$0.018$~\cite{:2008sq,pdg2008}&
 &
\footnotesize
\\
\end{tabular}
\end{ruledtabular}
\end{table*}
\endgroup

\begingroup
\begin{table*}
\caption{\label{table:data-dssl}
Summary of $D^+_s$ leptonic and semileptonic decays
used to perform the spectrum extrapolation.
Assumed branching fractions are shown in the second column;
normalization of each component is allowed to float within the
given uncertainty during the fit.
Form-factor models used to describe the
shape of each spectrum are shown in the third column:
single-pole (SPOLE~\cite{Becirevic:1999kt}) and
ISGW2~\cite{Scora:1995ty}.
The size of the expected secondary electron contribution
from the leptonic decay $D^+_s \to \tau^+ \nu_\tau$ is shown in the last row
based on the known branching fraction of
$D^+_s \to \tau^+ \nu_\tau$ decay~\cite{Alexander:2009ux,Onyisi:2009th},
and the shape is obtained from the \textsc{evtgen}~\cite{evtgen} MC program.
}
\begin{ruledtabular}
\begin{tabular}{llll}
Channel& $\mathcal{B}$ (\%)&
Form factor& Comment\\
\hline
$D^+_s \to \phi e^+ \nu_e$&
$2.36(26)$~\cite{Ecklund:2009fi}&
SPOLE &
$m_V=2.1$ GeV,
%
$m_A=2.28$ GeV,
\\
 &
 &
 &
$r_V=1.849(112)$,
and
$r_2=0.763(96)$
\\
 &
 &
 &
from Ref.~\cite{Aubert:2008rs}
\\
$D^+_s \to \eta e^+ \nu_e$&
$2.48(32)$~\cite{:2009cm}&
ISGW2 &

\\
$D^+_s \to \eta^\prime e^+ \nu_e$&
$0.91(33)$~\cite{:2009cm}&
ISGW2 &

\\
$D^+_s \to K^0 e^+ \nu_e$&
$0.37(10)$~\cite{:2009cm}&
ISGW2 &

\\
$D^+_s \to K^{\ast 0} e^+ \nu_e$&
$0.18(7)$~\cite{:2009cm} &
ISGW2 &

\\
$D^+_s \to f_0 e^+ \nu_e$&
$0.40(6)$~\cite{Ecklund:2009fi,Ablikim:2005kp} &
SPOLE &
%
$m_\text{pole} = 1.7$ GeV~\cite{Ecklund:2009fi}
\\
%
$D^+_s \to \tau^+ \nu_\tau$, $\tau^+ \to e^+ \nu_e \bar{\nu}_\tau$&
$1.003(79)$~\cite{Alexander:2009ux,Onyisi:2009th,pdg2008} &
 &
\\
%
\end{tabular}
\end{ruledtabular}
\end{table*}
\endgroup

\end{document}